\documentclass[useAMS,usenatbib,usegraphicx,a4]{mn2e}
\usepackage{times}
\usepackage{url}
\usepackage{epsfig}
\usepackage[flushleft]{threeparttable}
\usepackage{amssymb}
\usepackage{amsmath}
\usepackage{pdfpages}
\usepackage{todonotes}

\def\arcsec{^{\prime\prime}}

\def\apjl{ApJL}
\def\apj{ApJ}
\def\apjs{ApJS}
\def\mnras{MNRAS}
\def\araa{ARAA}
\def\aj{AJ}
\def\aap{A\&A}
\def\aaps{A\&A Suppl.}
\def\nat{Nature}
\def\pasp{PASP}

\newcommand{\refbf}{} 

\title[Extinction Trends with Inclination]{OMEGA -- OSIRIS Mapping of Emission-line Galaxies in A901/2: IV. -- Extinction of Star-Formation Estimators with Inclination}

\author[C. Wolf et. al.]{Christian Wolf$^{1,2}$, Tim Weinzirl$^{3}$, Alfonso~Arag\'on-Salamanca$^{2,3}$, Meghan~E.~Gray$^{3}$, 
\newauthor Bruno~Rodr\'iguez~del~Pino$^{4}$, Ana~L.~Chies-Santos$^{5}$, Steven~P.~Bamford$^{3}$,
\newauthor Asmus B\"ohm$^{6}$, Katherine Harborne$^{3,7}$ \\
$^{1}$Research School of Astronomy and Astrophysics, Australian National University, Canberra ACT 2611, Australia, E-mail: christian.wolf@anu.edu.au\\
$^{2}$Australian Research Council (ARC) Centre of Excellence in All-Sky Astrophysics (CAASTRO) \\
$^{3}$School of Physics and Astronomy, The University of Nottingham, University Park, Nottingham, NG7 2RD, UK\\
$^{4}$Centro de Astrobiolog\'ia, INTA-CSIC, Madrid, Spain\\
$^{5}$Departamento de Astronomia, Instituto de Física, Universidade Federal do Rio Grande do Sul, Porto Alegre, R.S, Brazil\\
$^{6}$Department of Astrophysics, University of Vienna, T\"urkenschanzstr. 17, 1180 Vienna, Austria\\
$^{7}$International Centre for Radio Astronomy (ICRAR), The University of Western Australia, Crawley, WA 6009, Australia\\
}

\begin{document}

\date{draft \today}
\maketitle


\begin{abstract}
We study the effect of inclination on the apparent brightness of star-forming galaxies in spectral passbands that are commonly used as star-formation indicators. As diagnostics we use mass-to-light ratios in three passbands: the UV continuum at 280~nm, the H$\alpha$ emission line, and the FIR 24$\mu$-band. We include a study of inclination trends in the IR/UV ratio (``IRX'') and the IR/H$\alpha$ ratio. Our sample comprises a few hundred galaxies from the region around the clusters Abell 901/902 with deep data and inclinations measured from outer disks in {\it Hubble Space Telescope} images. As a novelty, the H$\alpha$- and separately the NII-emission are measured by tunable-filter imaging and encompass galaxies in their entirety. At galaxy stellar masses above $\log M_*/M_\odot \ga 10$ we find trends in the UV and H$\alpha$ mass-to-light ratio that suggest an inclination-induced attenuation from face-on to edge-on of $\sim 1$ mag and $\sim 0.7$ mag in UV and H$\alpha$, respectively, {\refbf implying that star-formation rates of edge-on galaxies would be underestimated by $\sim 2.5\times$ in UV and $\sim 2\times$ in H$\alpha$}. We find the luminosities in UV and H$\alpha$ to be well correlated, but the optical depth of diffuse dust that causes inclination dependence appears to be lower for stars emitting at 280~nm than for gas clouds emitting Balmer lines. For galaxies with $\log M_*/M_\odot \la 9.7$, we find no measurable effect at $>0.1$~mag. The absence of an inclination dependence at 24$\mu$ confirms that the average galaxy is optically thin in the FIR. 
\end{abstract}
\begin{keywords}
galaxies: general -- galaxies: clusters: general -- galaxies: star formation -- dust, extinction -- infrared: galaxies -- ultraviolet: galaxies
\end{keywords}


\section{Introduction}\label{intro}

The appearance of galaxies is largely determined by their luminous components, i.e. stars and ionised gas clouds. A second factor is extinction of galaxy light due to dust inside them, where geometry and amount of dust determine how a galaxy appears from different viewing angles. Star-forming disk galaxies are the most common type of galaxy, and they contain a fair amount of dust, mostly in a disk thinner than the stellar disk \citep{Xilouris99}, at least in galaxies that are large enough to be ordered by significant rotation \citep{Dalcanton04}.

Disk galaxies viewed edge-on usually display impressive dust bands that clearly obscure much of a galaxy's stars and gas clouds, while those seen face-on show their dust more as a filigree pattern woven subtly into their spiral arms. This difference in prominence and optical depth of the dust is even clearer when considering viewing points inside the dust disk, e.g. the Earth's location inside the Milky Way. When we look out of our Galaxy's disk at steep angles, only a few percent of the light is absorbed before arriving at Earth. But when we look along a line-of-sight running within the disk itself, the view is quickly blocked; visual light from the centre of the Milky Way arrives at Earth only after being diminished by a factor of up to one trillion \citep{Catchpole90}. For a while, a debate has raged about whether disk galaxies are optically thin or thick \citep{Holmberg58,Disney89,Valentijn90,Xilouris99,Holwerda05}, but it seems there is convergence to the view that disks are optically thick only when viewed edge-on or in their innermost parts, see \citet{Calzetti01} for a review, and also \citet{Wild11}.

Many details of the spatial distribution between luminous components and absorbing dust are important for the appearance of a galaxy in different spectral passbands, as discussed e.g. by \citet{Popescu11}. However, the viewing angle of a galaxy, or inclination, is still an obvious primary ordering parameter that can be easily measured. Hence, several past works have considered the effect of inclination on the apparent sizes, concentration, brightness and colour of galaxies \citep[e.g.][]{Holmberg58, Giovanelli94, Tully98, Masters03, Driver07, ChoPark09, Maller09, M10, Wild11, Devour16, Devour17, Battisti17}. It is found that the inclination of a galaxy affects estimates of its physical properties, such as the population mix and total mass of its stars \citep[see review by][]{Conroy13}; when considering samples of galaxies it causes an underestimation of the average cosmic matter density \citep[e.g.][]{Driver07}; 
it complicates attempts to understand galaxy evolution from comparisons with simulations \citep[e.g.][]{SomPrim99, McKinnon17}; and it causes apparently random, flux-limited samples of galaxies to suffer from selective incompleteness that even affects cosmological studies; finally, measurements of the star-formation rate are affected by inclination \citep{Morselli16,Leslie18}. 

On the other hand, changes in galaxy appearance with inclination are an opportunity to learn about the dust itself. A first step towards a more structured picture of dust is a two-component model \citep[e.g.][]{LonsdalePerssonHelou87,CF00,Tuffs04} with diffuse dust and clumpy dust: diffuse cirrus-like dust pervades the disk of the Milky Way and other galaxies with a high fill factor at low density, while high-density dust clumps enshroud the birth clouds of stars with high optical depth but moderate and constantly evolving covering factors as the radiation from the new-born stars blows away the dust within less than ten million years. 

Most observations to date appear consistent with this two-component picture \citep[e.g.][]{Wild11}, including the fact that Balmer emission from star-forming regions is more extinguished than the stellar continuum of older stars by a factor of two or more \citep{Kennicutt83,Calzetti94,Yip10,Hao11,Wild11} due to high optical depth of the dusty birth clouds. When galaxies are inclined in the two-component model, the line-of-sight through cirrus becomes longer and the optical depth of dust larger; however, the total extinction of new-born stars embedded in birth clouds increases less in relative terms than the extinction of older stars, whose face-on extinction by the low-density cirrus is expected to be mild. As the extinction of stars depends not only on inclination but also age, the effective attenuation curves of the integrated stellar population in a galaxy may change with inclination. Effective attenuation curves are light-weighted and thus susceptible to changes in the apparent mix of stellar populations in the integrated light of a galaxy, even when the local composition and extinction law of the dust is constant throughout the galaxy \citep{Calzetti94,Wild11,Battisti17}. 

Previous studies have shown that dust opacity in galaxy disks has (i) a radial dependence, where centres are more extinguished than outskirts of disks \citep{Valentijn94,Peletier95,Holwerda05,Tacchella17}; (ii) a type dependence, where early-type spirals have higher opacities than late-type disks \citep{deVau91,Han92,M10}; (iii) a luminosity dependence, where the mean extinction of the disk increases towards more luminous galaxies \citep{Giovanelli95,Tully98,Masters03}, but turns over at the highest luminosities, where the specific star formation rate drops \citep{M10,Devour16}; and (iv) spiral arms are more opaque than inter-arm regions \citep{Beckman96, White00, Holwerda05}. \citet{ChoPark09} found that extinction changes with concentration of the galaxy, and \citet{Grootes13} note that the central face-on $B$-band dust optical depth in a galaxy is correlated with its mean stellar mass surface density.

Previous work has mostly addressed the effects of inclination on the appearance of galaxies in the stellar continuum light of UV-optical and near-infrared passbands. Generally, it is observed that measuring additional extinction as galaxy orientation changes from face-on to edge-on is easier than assessing total extinction, although approaches exist for the latter \citep[e.g.][]{Driver07}. For instance, \citet{M10} find that the average extinction added from face-on to edge-on in the passbands of Sloan Digital Sky Survey \citep[SDSS,][]{SDSS} ranges from 0.4~{\refbf mag} in $i$-band to 0.7~{\refbf mag} in $u$-band for the integrated light of a galaxy, and \citet{Devour16} find similar values but can distinguish trends with mass and star formation. Studies of integrated light of course average over a range of behaviour from negligible extinction in the outer parts to high obscuration in galaxy centres. Some work has considered trends in the extinction of Balmer lines, but such data was mostly drawn from SDSS fibre spectra, which concentrate on small central areas in galaxies and thus fail to represent their disks \citep[e.g.][]{Yip10,Wild11,Battisti17}.

In this paper, we look specifically at the inclination-dependence of total galaxy luminosity in passbands that are used as star-formation indicators, notably the UV~280~nm continuum, the integrated H$\alpha$-line luminosity and the far-infrared continuum in the {\it Spitzer} 24$\mu$ band. 
Specifically, the {\refbf 24$\mu$} luminosity is a star-formation indicator that should be largely independent of inclination, such that the ratio of UV-to-IR and H$\alpha$-to-IR luminosity can {\refbf directly reflect UV- and H$\alpha$-extinction}. 

We work with a comparatively small sample of a few hundred galaxies, but compared to SDSS {\refbf our spatial resolution is larger, and also our multi-band images have vastly deeper surface brightness sensitivity; this allows fitting the outer disk contours as opposed to relying on inner axis ratios that are more easily biased by bulges and bars; finally, our deep H$\alpha$ and [N{\sc ii}] line imaging captures H$\alpha$ light from across the entire galaxy disk, thus matching the H$\alpha$-footprint consistently to the other passbands and avoiding aperture biases}. We use a volume-limited sample of  galaxies around a cluster at a fixed distance (redshift $z\approx 0.16$) and thus avoid a dispersion of K corrections within the sample. 


In Section~\ref{s-data-sample}, we describe our multi-wavelength data and the sample, including inclination biases in stellar-mass determination. In Section~\ref{results} we measure the slopes of our $M_*/L_{\rm band}$ ratios in UV, H$\alpha$ and FIR versus stellar mass and inclination, and consider in particular the infrared excess over UV and H$\alpha$. This is followed by a discussion and summary. We adopt a flat $\Lambda$ cold dark matter cosmology with $\Omega_\Lambda=0.7$ and $H_0=70$~km~s$^{-1}$~Mpc$^{-1}$, which fixes the luminosity distance to our target cluster at $D_L=791$~Mpc. We use Vega magnitudes throughout the paper, and a bolometric luminosity for the Sun of $L_{\odot,\rm bol} = 3.823 \times 10^{33}$~erg~s$^{-1}$.

\section{Data}\label{s-data-sample}

This work brings together data from several wavelengths and observatories, while it focuses on one particular galaxy cluster system at $z\sim 0.165$, the Abell 901/2 system of four sub-clusters. We aim to study how the brightness of star-forming disk galaxies is affected by inclination, and thus combine measurements of galaxy inclination and stellar mass with measures of mass-to-light ($M_*/L$) ratios in different spectral passbands. Our interest lies primarily in pass-bands used as star-formation rate indicators. We include: 
\begin{enumerate}
\item luminosities observed in the UV continuum at rest-frame 280~nm, which is emitted mostly by young stars of $<100$~Myr age and which may have migrated away from their birth regions but still reside in the thin star-forming disk of a galaxy; 
\item the H$\alpha$-emission line, which mostly originates from star-forming regions with very young stars of $<10$~Myr age surrounded by ionised gas (HII regions), but includes a significant component of diffuse interstellar gas; \citet{Popescu11} suggest that 60\% of UV photons escape the clumpy dust around HII regions, and \citet{Calzetti13} suggest that these could ionise gas within 1~kpc of the star-forming region; 
\item the far-infrared continuum at $24\mu$, which is emission from dust heated mostly by young stars, but partially by intermediate-age stars \citep[and references therein]{Calzetti13}; by mass most of the dust {\refbf is} in a diffuse component, while a fraction of perhaps $<15$\% is clumpy and enshrouds star-forming regions \citep{Grootes13}. 
\end{enumerate}

We restrict the analysis to normally star-forming galaxies, where these measurements are not affected by light from active galactic nuclei (AGN). We identify and deselect AGN using their [N{\sc ii}]/H$\alpha$ line ratio, and we separate dust-poor red-sequence galaxies from normally dusty star-forming galaxies {\refbf with}  optical SEDs.

\subsection{Data sources}\label{data-data}

This cluster system was initially observed in detail by the survey {\it Classifying Objects by Medium-Band Observations in 17 bands} \citep[COMBO-17,][]{W03,W04}, which provided photometric redshifts for most of the cluster galaxies with a precision of $\delta z <0.005$, on the order of the velocity dispersion of the individual sub-clusters and smaller than some of the velocity differences between them. From this data, \citet{WGM05} selected a largely complete sample of 795 cluster members down to a luminosity of $M_V= -17$. COMBO-17 provides luminosities for several passbands obtained by SED fitting. The COMBO-17 SEDs are corrected for foreground extinction by interstellar dust ($A_R=0.1476$). {\refbf Measurements of the rest-frame 280~nm passband at $z\sim 0.16$ required} minor extrapolation as the bluest observed filter is centred on a rest frame wavelength of 310~nm; this is expected to change the estimates of $L_{280}$ by less than 1--2\%. All objects in the cluster sample have reliably measured rest-frame 280~nm luminosities, as the sample cut does not exploit the depth of the COMBO-17 imaging. 

\citet{WGM05} also presented a colour-colour diagram to isolate the true red sequence from star-forming galaxies across their whole range of colour, and showed how in some colour indices star-forming galaxies can be even redder than quiescent galaxies with old populations \citep[see also][]{M10}. As a result, they distinguish between quiescent and dust-poor {\it old red} galaxies and similarly red but moderately star-forming and moderately dusty galaxies, which they called {\it dusty red} to contrast them against truly old red cases. In this work, old red galaxies are removed from the sample as they are expected to have little dust and star formation. 


{\refbf Observations with {\it Spitzer} allowed the study of obscured star formation in the cluster galaxies from MIPS 24$\mu$ data \citep{B07,Gallazzi09,W09}. Sources are detected to a flux limit of 58~mJy, corresponding to a star-formation rate of $0.14 M_\odot/$yr \citep{B07}, but} we expect the {\refbf sample} to be only $\sim30$\% complete at this level, while 80\% completeness is reached at 93~mJy. Given a PSF of $\sim 6\arcsec$ FWHM, or 18~kpc at the cluster distance, galaxies are mostly unresolved in the 24$\mu$ images, but are thus also not susceptible to size and inclination biases in detection. 

\citet{B04} estimated stellar masses for COMBO-17 galaxies using the prescription of \citet{BdJ01}. Based on that the cluster sample is complete at $\log M_*/M_\odot \ga 8.75$ for a \citet{Chabrier03} initial mass function (IMF). The COMBO-17 SEDs underlying these mass estimates are, however, constructed from central-aperture ($\sim 4.5$~kpc) SED measurements that are corrected for aperture losses using a single deep ($R_{5\sigma}=26$) $R$-band image from COMBO-17. This implies that the mass-to-light ratios are biased in large galaxies with colour gradients; specifically, they are probably overestimated by $\la 0.25$~dex in spiral galaxies of $\log M_*/M_\odot >11$ \citep{W09}. In this paper, we use stellar mass itself first as an ordering parameter for the galaxy population, where a stretch of the scale at the massive end is inconsequential for our analysis. The H$\alpha$-selection explained below also eliminates the most massive galaxies that are most affected by this problem from the analysis. We also use mass-to-light ratios as a crucial diagnostic for extinction, so that biases in their derivation are important, and we will discuss these in Sect.~\ref{data-mlr}.

The field was then observed with the {\it Hubble Space Telescope} (HST) in the F606W filter, and covered by 80 tiles of the {\it Advanced Camera for Surveys} (ACS), as part of the {\it Space Telescope Abell 901/2 Galaxy Evolution Survey} \citep[STAGES,][]{Gray09}. All galaxies in the cluster sample are well-detected in the $\ga 2,000$~sec HST exposures and show light profiles to a surface brightness of $\mu_{\rm V606}=26.5$~${\rm mag~arcsec^{-2}}$. The superior spatial resolution of $\sim 0.1\arcsec$ supplies morphological detail at a resolution of 300~pc. {\refbf In this work the HST data are used} to measure disk inclinations and apertures to re-derive multi-band photometry.

{\refbf More recent observations of the H$\alpha$ and [N{\sc ii}] emission lines were} obtained by the {\it OSIRIS Mapping of Emission-line Galaxies in A901/2} survey \citep[OMEGA,][]{OMEGA1,OMEGA2}. {\refbf These were carried out with the {\it Optical System for Imaging and low Resolution Integrated Spectroscopy} (OSIRIS) instrument \citep{Cepa13} on the 10.4~m Gran Telescopio Canarias (GTC) and cover $\sim 70$\% of the COMBO-17 area}. The data set consists of Fabry-P\'erot narrow-band images with 14~\AA\ FWHM, from which low-resolution emission-line spectra were constructed {\refbf with up to 48 wavelength points} for each galaxy. Spectral fluxes are measured in {\refbf galaxy-specific apertures of} 2.5$\times$ the size of the galaxy semi-major axis in the HST/F606W image. Within the cluster sample defined above, OMEGA detected 439 galaxies with an H$\alpha$ flux of $f_{\rm H\alpha} >3\times 10^{-17}$~erg/s/cm$^2$ and an H$\alpha$ equivalent width of $W_{{\rm H}\alpha} >3$~\AA\ \citep{OMEGA2}. The separately measured [N{\sc ii}]-line allows us to type galaxies into star-forming vs. AGNs. We note that we have no data on the H$\beta$-line, which is often used to estimate extinction on Balmer line-emitting gas.

{\refbf Since the compilation of \citet{Gray09}, updates have been made to the Spitzer data set (adding 14 objects by improved cross-matching) and the OMEGA data were only obtained later. We also rederived improved total UV luminosities to match the total fluxes measured in the 24$\mu$ and H$\alpha$ images}.


\subsection{Galaxy inclinations}\label{data-inc}

Galaxies are inclined at mostly random angles in the sky, but their inclination angle may affect their brightness depending on the passband of observation. While stars are expected to radiate mostly isotropically, {\refbf they are} viewed through lines-of-sight traversing different parts of their galaxy. As a galaxy's inclination dictates the column density of dust along that line-of-sight, it determines how much light is absorbed from young stars and star-forming regions. Large edge-on galaxies, e.g., are seen through the longest line-of-sight of extinguishing dust and typically appear very red in UV-optical colours, irrespective of their star-formation rate.

We define galaxy inclination $i$ to be the angle between a galaxy's disk plane and the plane of the sky, such that face-on galaxies have $i=0^\circ$. Assuming an axisymmetric, infinitesimally thin disk, the inclination angle is related to the ratio of semi-minor axis to semi-major axis ($b/a$) by $\cos i = b/a$, but \citet{Hubble26} already accounted for the thickness of a galaxy from an intrinsic axis ratio $q$ using the equation 
\begin{equation}
\cos i = \sqrt{\frac{(b/a)^2-q^2}{1-q^2}}~. 
\end{equation}

We follow the literature, which has long used $\log a/b$ as an ordering parameter for inclination \citep[e.g][]{deVau91, Giovanelli94, M10}, and has measured dust extinction by fitting the slope $\gamma$ in the expression $A_{\rm band} = a + \gamma \log a/b$, where $A_{\rm band}$ is the total attenuation of light in a passband. We take the intrinsic thickness of galaxy disks into account by capping the axis ratio at $\log a/b = 0.8$, equivalent to an average axis ratio of edge-on galaxies of $q \approx 0.16$ \citep[similar to][]{Maller09}, while measured values range from 0.1 to 0.22 for Sd to Sa galaxies \citep{UR08,M10}.

A standard way to measure galaxy inclination is to fit ellipses to a photometric image \citep[e.g.][]{Jed87}. We use the \texttt{IRAF} $ellipse$ task with a fixed centre to determine the ellipticity and position angle of the outer galaxy disc at a surface brightness of $\mu_{V606,\rm AB}=26.5$~${\rm mag~arcsec^{-2}}$ \citep[for details see][]{Maltby12}. This approach minimises bias from bulges on the inclination measurement except for nearly edge-on viewing angles, where the bulge may appear wider than the outer disk itself. 



\begin{figure*}
\begin{center}
\begin{minipage}{0.385\textwidth}
 \includegraphics[angle=270,width=\columnwidth,clip=true]{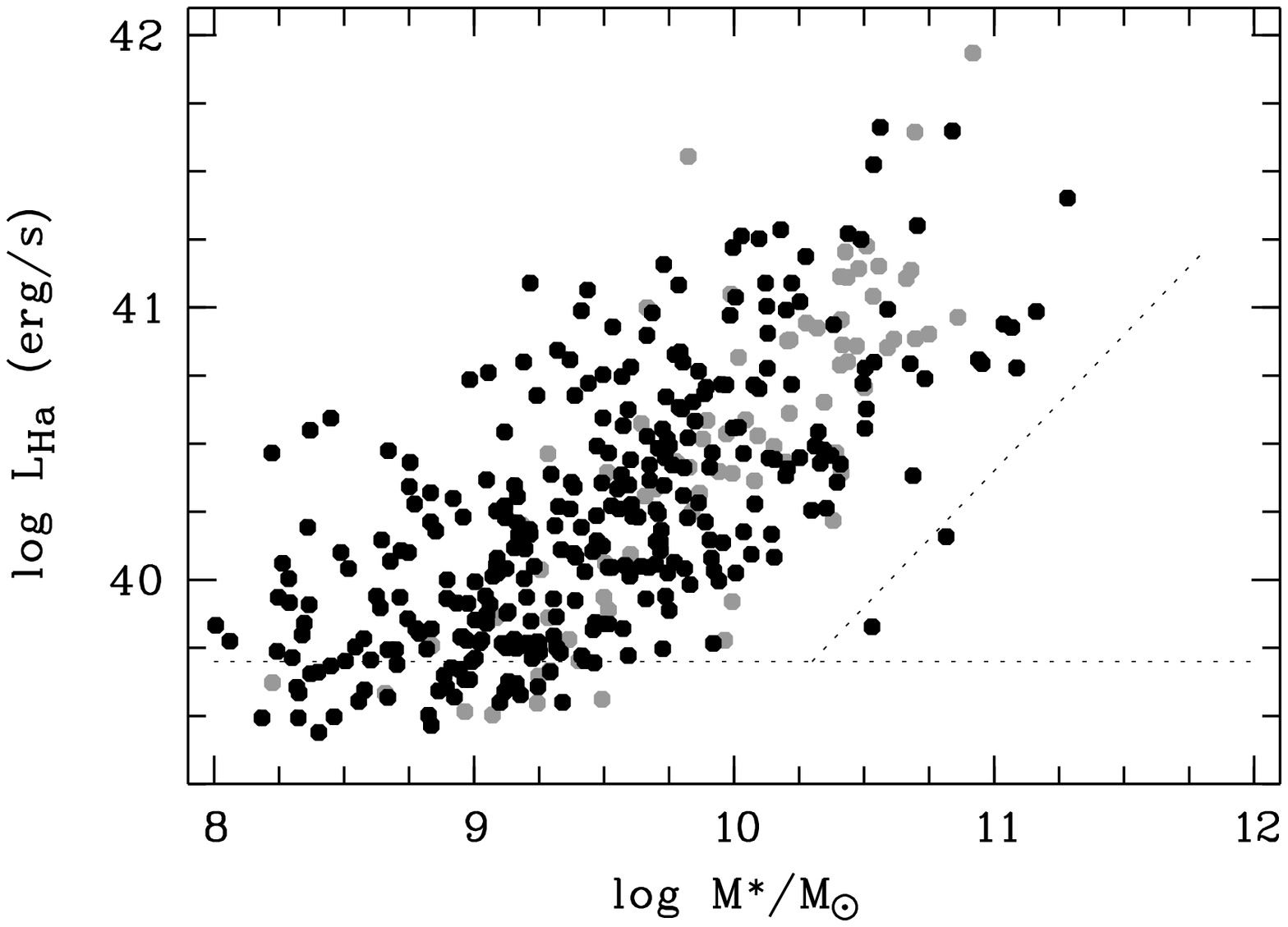} 
\end{minipage}
\hfill
\begin{minipage}{0.59\textwidth}
 \includegraphics[angle=270,width=\columnwidth,clip=true]{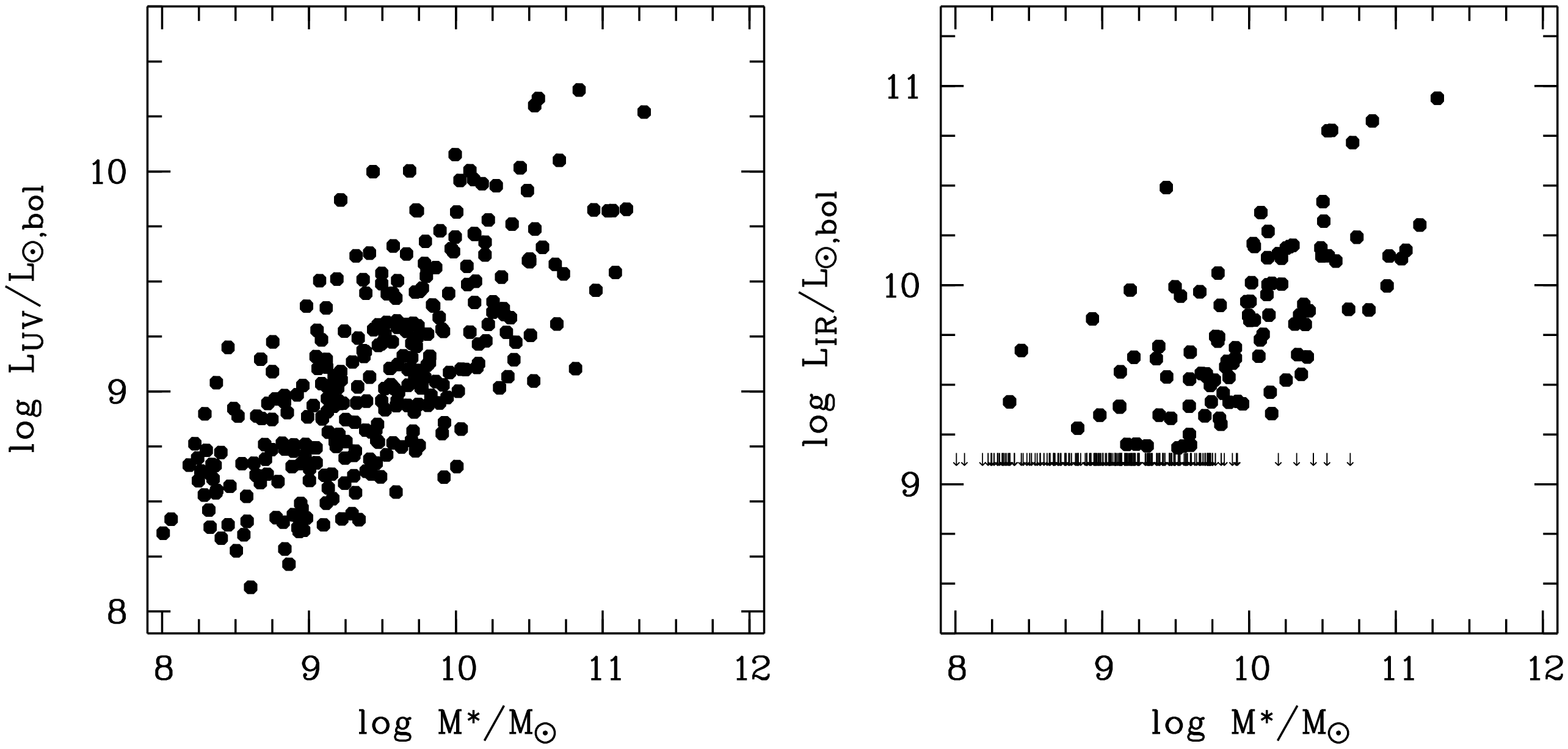} 
\end{minipage}
\caption{Sample of star-forming galaxies in the cluster Abell 901/2 used in this paper. {\it Left:} H$\alpha$-line luminosity vs. stellar mass; the main sample (dark points) is complete at line fluxes $>7\times10^{-17}$~erg$/$s$/$cm$^2$, corresponding to a luminosity of $\log L_{{\rm H}\alpha}/({\rm erg~s^{-1}}) = 39.72$ (horizontal line) at $D_L=791$~Mpc; it is also complete at $W_{{\rm H}\alpha}>3$~\AA\ (diagonal line). Light symbols mark objects removed from the sample because of AGN or shock contribution to their emission lines. {\it Centre:} UV 280~nm luminosity of the main sample. {\it Right:} IR luminosity from {\it Spitzer} data with detection limits at $\log L_{\rm IR}/L_{\odot,\rm bol} = 9.16$.  
\label{sample}}
\end{center}
\end{figure*}

\subsection{Sample definitions}

In this work, we consider galaxy luminosities and especially mass-to-light ratios in three different passbands as a function of galaxy mass and inclination. {\refbf However,} the detections in these passbands come with different limits and are made in different footprints. The data for the UV continuum at 280~nm cover the widest area and has the deepest detection limits. The H$\alpha$ data covers $\sim 70$\% of that area and have detection limits such that 439 out of 795 galaxies have both UV and H$\alpha$ detections. 

Fig.~\ref{sample} shows this sample in the left panel, together with two lines approximating completeness limits: the horizontal line at an H$\alpha$ luminosity of $\log L_{{\rm H}\alpha}/({\rm erg~s^{-1}}) = 39.72$ represents a completeness limit corresponding to a line flux of $7\times10^{-17}$~erg$/$s$/$cm$^2$ at the luminosity distance of A901/2 ($D_L=791$~Mpc), below which the number counts of H$\alpha$-detections turn over. We cut the sample at an H$\alpha$ equivalent width of $W_{{\rm H}\alpha} = 3$~\AA , approximated by a diagonal line, below which the line detection itself is unreliable. This cut effectively excludes only red-sequence galaxies with little star formation. 

From the H$\alpha$ sample of 439 galaxies we reject 47 objects that are either flagged with large errors or are classed as red-sequence galaxies ('old red') based on the COMBO-17 SED, which means that their H$\alpha$ and UV emission is unlikely to originate from young stars and star formation. Next we reject 77 galaxies, whose [N{\sc ii}]/H$\alpha$ line ratios suggest that they are likely affected by either nuclear activity or environmentally induced shocks, following the WHAN diagram of \citet{CidF11}. We identify these objects on the basis of the OMEGA line-fit, and consider galaxies as possible AGN when the Bayesian fit probability is over 90\% that $\log [{\rm N}{\textsc {ii}}]/{\rm H}\alpha >-0.4$ \citep[see][]{OMEGA2}; the likely AGN mostly show lower specific H$\alpha$ luminosity. This leaves 315 galaxies in the combined UV/H$\alpha$ sample; the middle panel of Fig.~\ref{sample} shows their UV luminosities.

Out of these 315 galaxies, 297 are contained in the footprint of the 24$\mu$ data (see right panel of Fig.~\ref{sample}); this includes 109 detections above 58~mJy arranged in a star-forming main sequence of galaxies, that appears to be complete at $\log M_*/M_\odot \ga 9.75$, where we find 74 objects, with the most massive galaxy having an estimated mass of $\log M_*/M_\odot =11.28$. In the following, we use two samples:
\begin{itemize}
\item an IR-complete higher-mass sample with 74 galaxies selected by $\log M_*/M_\odot > 9.75$ and $\log L_{\rm IR}/L_{\odot,\rm bol} > 9.16$ that roughly corresponds to ${\rm SFR_{IR}} \ga 0.14 M_\odot /$yr; and 
\item a UV-complete lower-mass sample  with 172 galaxies selected by $\log M_*/M_\odot =[8.75,9.75]$, drawn from a slightly larger area.
\end{itemize}

We note that inclination-dependent incompleteness of our sample should not affect the isotropic IR- and deep UV-data, but they could affect H$\alpha$ detections of highly inclined galaxies at the lower-mass sample limits; however, previous works have found only little overall extinction in low-mass galaxies \citep{Tully98,M10}, so this effect is not expected to be significant.

\subsection{Mass-to-light ratios and star formation}\label{data-mlr}

Most of the analysis in this paper concerns trends of dust extinction with galaxy inclination, which could be inferred {\refbf in two ways: first by measuring inclination  trends of luminosity in an extinguished passband relative to a not extinguished one, and second by measuring inclination trends in the mass-to-light ratio of galaxies. We will first do the latter, although it is less reliable, because we can apply it to both the low-mass and the high-mass sample; then we come back to the former, which is the better measure, but for us works only in the high-mass sample. }

Using a sample at fixed luminosity distance and ignoring a mean $M_*/L_{\rm band}$ ratio for face-on galaxies, $M/L$-trends can be measured from (i) apparent total fluxes in the relevant passbands and (b) stellar masses. Total fluxes are moderately straight-forward to measure, although inclination can affect surface brightness and thus sensitivity to total light. {\refbf We avoid biases from surface brightness changes by using consistent, large apertures in H$\alpha$ and UV (the unresolved 24$\mu$ images are independent of inclination). }

The stellar mass, however, is derived from an estimate of the $M_*/L_V$ ratio that is based on the dust-extinguished optical SED from the inclined galaxy. We estimate stellar mass using the method of \citet{BdJ01}, although the reliability of their approach has been debated: for dust extinction with $R_V=3$, their masses are supposed to be correct in an optically thin extinction scenario, where the loss of light is exactly compensated by an increase in the inferred $M_*/L_V$ ratio due to reddening. However, as the colour of a galaxy is light-weighted, so is the $M_*/L_V$ ratio, and heavily absorbed galaxy components behind optically thick dust will fail to be taken into account in both light and mass. 

Hence, if the method for estimating $M_*/L$ from the SED has some inclination dependence that is not accounted for, such a bias trend with inclination will propagate into the trends we derive. Empirically, \citet{Driver07} argue that \citet{BdJ01} masses are fine up to $i \le 60\degr$, but underestimated towards edge-on, by up to $\Delta \log M \approx 0.2$ or 0.4~dex for disks and bulges, respectively, based on their Fig.~13. \citet{Maller09} compare mass histograms of face-on and edge-on galaxies and find that \citet{BdJ01} masses are the least biased among a choice of three varieties; while they tend to be slightly underestimated for edge-on galaxies, the difference in their histograms is well below 0.1~dex. 

\begin{figure}
\begin{center}
\includegraphics[angle=270,width=\columnwidth,clip=true]{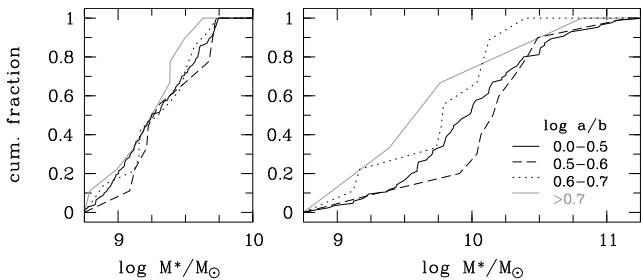} 
\caption{Cumulative stellar-mass histograms of galaxies in inclination bins for the low-mass sample (left) and the all IR-detected galaxies with $\log L_{\rm IR}/L_{\odot, \rm bol} > 9.16$ (right). The IR-detected sample shows signs of mass underestimation at higher inclination.
\label{masshisto}}
\end{center}
\end{figure}

Our first attempt to estimate mass biases at high inclination uses cumulative stellar-mass histograms of sub-samples in bins of inclination. For a randomly oriented sample of galaxies, these histograms should be similar if there is no inclination bias in the mass determination. Fig.~\ref{masshisto} shows mass histograms of our lower-mass sample and of the full IR-detected galaxy sample that we expect to radiate approximately isotropically; for the full IR-sample we drop the mass cut that makes the higher-mass sample so we can see if we lose objects due to underestimated mass. We combine inclinations up to $\sim 71\degr$ or $\log a/b < 0.5$ into a single bin (solid line) after not finding trends when subdividing into further bins. 
There is a clear signature that galaxies at $\log a/b > 0.7$ ($i \ga 80\degr$) have underestimated masses, while in the range of $\log a/b = [0.6,0.7]$ the situation is less clear. 
We proceed with two versions of the sample, one with inclinations limited to $\log a/b < 0.6$ and one without limits. The limit reduces the higher-mass sample from 74 to 67 galaxies, and the lower-mass sample from 172 to 149 galaxies. 

It is also possible to search for inclination biases by including in our analysis a passband in which we expect galaxies to radiate isotropically. Any inclination bias in determining $M_*/L_V$ would manifest with full strength as an inclination trend in the $M_*/L_{\rm band}$ ratio of such a passband. Here, we use the far-infrared 24$\mu$ band, as we expect the dust emission from the galaxy disk to be optically thin and intrinsically isotropic. 

Finally, with 24$\mu$ emission likely to be optically thin or not extinguished, we consider trends in UV-to-IR and H$\alpha$-to-IR ratios to eliminate mass biases entirely from the inclination trends, and get the most reliable trends in UV- and H$\alpha$-extinction, but only for the high-mass sample.

\begin{figure*}
\begin{center}
\includegraphics[angle=270,width=0.95\textwidth,clip=true]{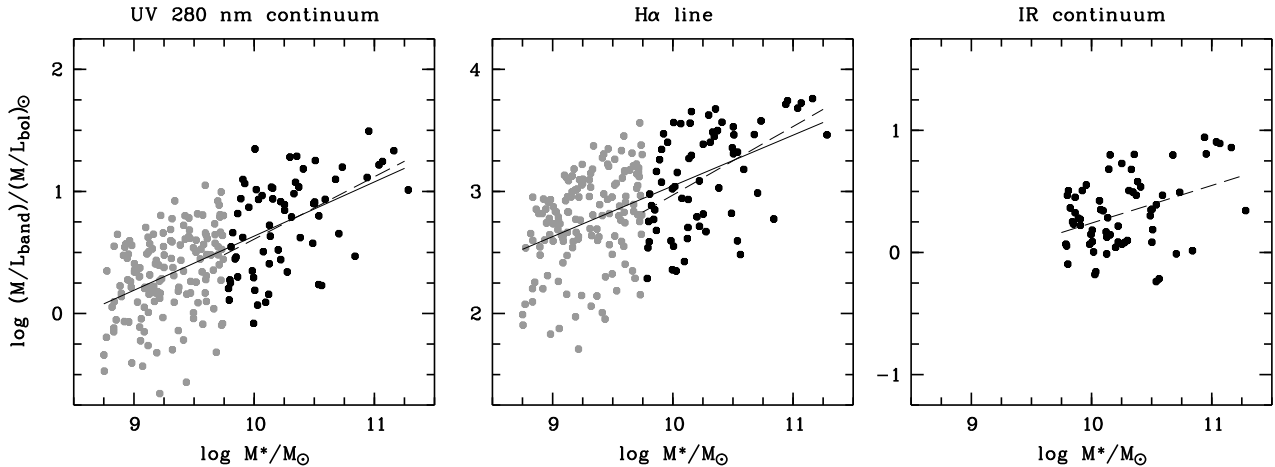} 
\caption{Trends of mass-to-light ratio with galaxy stellar mass in three bands: UV 280~nm continuum (left), H$\alpha$-emission line (centre) and IR 24$\mu$ (right). Dark symbols represent the IR-complete high-mass sample with $\log M_*/M_\odot =[9.75,11.25]$ and $\log L_{\rm IR}/L_{\odot, \rm bol} > 9.16$, which corresponds roughly to SFR~$> 0.14$~$M_\odot/$yr; light symbols represent the H$\alpha$- and UV-complete low-mass sample with $\log M_*/M_\odot =[8.75,9.75]$. Lines are regression fits to the combined samples (solid) and higher-mass-only samples (dashed).  
\label{masstrends}}
\end{center}
\end{figure*}

Several methods for estimating star-formation rates (SFR) assume a linear dependence on a passband luminosity, although there are non-linear SFR calibrations \citep[e.g.][]{Calzetti07,Davies16,Brown17}. {\refbf Linear calibrations have been proposed for all three passbands used in this paper \citep[for reviews see][]{KennEvans12,Calzetti13}. For linear estimators, the mass-to-light ratio is proportional to inverse specific star-formation rate. Hence, the ordinate axes in Figs.~\ref{masstrends} to \ref{incltrends_lowmass} could be labelled with SSFR$^{-1}_{\rm band}$ instead of $M_*/L_{\rm band}$. We use the latter as it makes no assumption on the SFR calibration. In either case, any trend of $M_*/L_{\rm band}$ ratio with inclination causes an equally strong trend in SFR estimates with inclination.} Our findings could thus be used to correct SFR estimates from these passbands, where galaxy inclination is known, which could help to tighten relationships between SFR and {\refbf other galaxy parameters}.

Assuming the {\refbf SFR calibrations previously used by \citet{B05,Gray09} and \citet{OMEGA2},} our sample has detection limits of $\rm (SFR_{UV}, SFR_{H\alpha}, SFR_{24\mu}) = (0.013,0.010,0.140)~M_\odot~yr^{-1}$ and 90\%-completeness limits of $\rm (0.020,0.023,0.25)~M_\odot~yr^{-1}$, although H$\alpha$ limits are higher for high-mass, low-EW galaxies.

\section{Results}\label{results}

In this section, we investigate the mass-to-light ratio of galaxies in the passbands UV 280~nm, H$\alpha$ and IR 24$\mu$ as a function of mass and inclination. The $M_*/L_{\rm band}$ ratios depend not only on inclination, but also on two further parameters, which add scatter to any trends with inclination:

\begin{enumerate}

\item There is a pronounced trend of $M_*/L_{\rm band}$ ratio with stellar mass of the galaxy. The mean colour and mean $M_*/L_{\rm band}$ ratio of star-forming galaxies is a function of stellar mass as a consequence of a mass trend in mean stellar age and specific star-formation rate \citep[the so-called star-forming main sequence, e.g][]{Noe07,Schim07,Wuyts11}. In addition, the mean dust column density of a galaxy may be a function of mass, which \citet{M10} have actually measured. In the following, we make no assumption on the relative importance of stellar population vs. dust factors for the trend, but return to this point in Sect.~\ref{discussion}. Hence, we first determine trends with mass and then try to control for them when investigating the inclination dependence of $M_*/L_{\rm band}$ ratio residuals relative to the mean at a given mass.

\item The $M_*/L_{\rm band}$ ratio also depends on the specific star-formation rate of an individual galaxy, i.e. its deviation from the star-forming main-sequence, which adds further scatter to $M_*/L_{\rm band}$ ratios in a galaxy sample. We will later suppress this source of scatter by investigating per-galaxy light-to-light ratios (a.k.a. colours) from pairs of passbands; these have no dependence on the star-formation rate itself, if the two passbands are both linear predictors of SFR; in this case, they convey instead only how light in different bandpasses is extinguished to a different degree depending on the inclination. 

\end{enumerate}

We remind the reader here that optical continuum passbands such as $ugriz$ are not linear predictors of star-formation rate, while the three passbands studied by us have all been used as linear SFR predictors before, despite certain limitations.

Overall, we model the observed $M_*/L_{\rm band}$ ratios in each passband $i$ as a linear relationship of the form:
\begin{equation}
  \log M_*/L_i = \alpha_i + \beta_i \log M_*/M_\odot + \gamma_i \log a/b  ~.
\end{equation}
We will obtain linear fits to observed passband luminosity ratios, $\log L_i/L_j$, with the same parametrisation, where $\beta_{i/j}$ quantifies the mass dependence and $\gamma_{i/j}$ the inclination dependence. 

\subsection{Mass dependence in mass-to-light ratios}

First, we investigate trends in the mass-to-light ratio vs. stellar mass for our three passbands (see Fig.~\ref{masstrends}) and in this case choose to eliminate high-inclination objects with $\log a/b > 0.6$ or $i\ga 75\degr$. In UV and H$\alpha$ we have two samples available for lower and higher masses, which we consider both independently and in combination, while the IR has only the higher-mass sample. We obtain linear regression fits and list the resulting slopes in Tab.~\ref{mass_slopes}. Using only a partial mass range leads to weaker constraints due to reduced leverage and makes the fits more influenced by outliers near the ends of the mass range. 

Using the largest possible sample in each case, we find slopes of $(\beta_{\rm UV},\beta_{\rm H\alpha},\beta_{\rm IR}) = (0.444, 0.415, 0.309)$ and $1\sigma$-uncertainties of 0.041, 0.045 and 0.092, respectively. Using only the higher-mass sample nearly triples the uncertainty in UV and H$\alpha$ to $\sim 0.12$ while making the slopes steeper. However, in Fig.~\ref{masstrends} we note six galaxies with relatively high $M_*/L_{\rm band}$ ratio at the highest masses, which increase the slope estimates in the higher-mass-only fit. At these high masses the star-forming main-sequence is sparsely populated, and these galaxies are instead probably already transitioning to the red sequence; hence, we prefer the combined fit. The purpose of this regression fit is to take out most of the nuisance variation in $M_*/L$ caused by mass in order to make the inclination study more sensitive. Hence, uncertainties in the slope $\beta$ won't appear as biases in the inclination study but only as a sub-dominant source of noise. 

\begin{table}
\centering
\caption{Slope values $\beta_i$ from linear fits to trends in mass-to-light ratio with stellar mass. The IR passband is only complete in the higher-mass sample.}
\label{mass_slopes} 
\begin{tabular}{l|cc|cc|}
\hline \noalign{\smallskip}  
		& \multicolumn{2}{|c|}{higher-mass sample}	& \multicolumn{2}{|c|}{combined sample}   \\
passband	& $\beta_i$ 		& rms 				& $\beta_i$ 		& rms 	    \\
\noalign{\smallskip} \hline \noalign{\smallskip}
UV 280		& $0.511 \pm 0.113$ & $0.34 $ & $0.444 \pm 0.041$ & $0.33$ \\
H$\alpha$		& $0.561 \pm 0.122$ & $0.37 $ & $0.415 \pm 0.045$ & $0.36$ \\
IR 24$\mu$	& $0.309 \pm 0.092$ & $0.27 $  \\
\noalign{\smallskip} \hline
\end{tabular}
\end{table}

\begin{figure*}
\begin{center}
\includegraphics[angle=270,width=0.95\textwidth,clip=true]{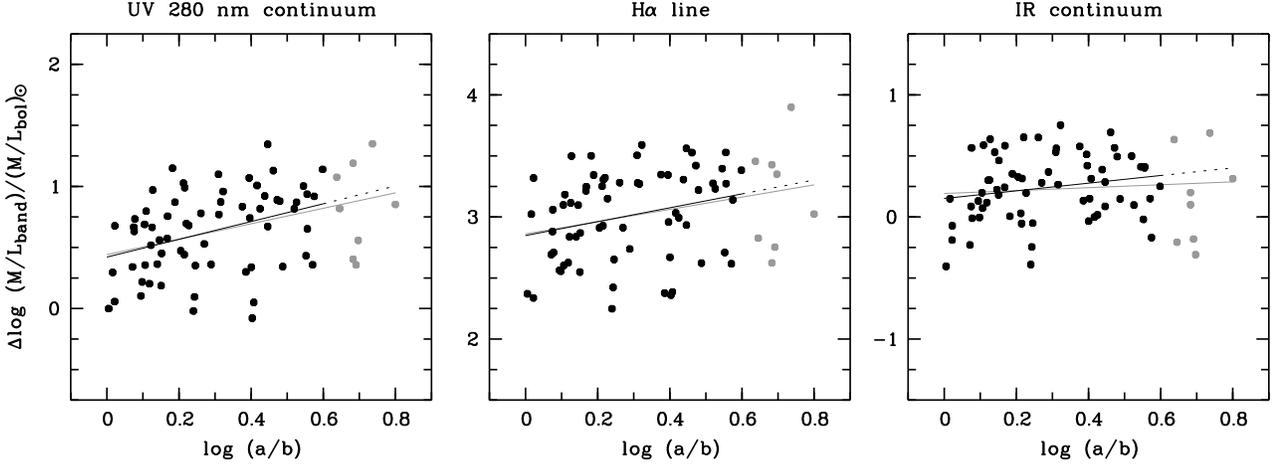} 
\caption{Trends of mass-to-light ratio with galaxy inclination in the high-mass sample of $\log M_*/M_\odot > 9.75$ and $\log L_{\rm IR}/L_{\odot, \rm bol} > 9.16$, after removing the mass trend: significant trends are seen in UV 280~nm continuum (left) and H$\alpha$-emission line (centre), but not in IR (right). Dark symbols and fits are inclination-restricted, while light symbols are highly inclined galaxies and light fits include all galaxies. 
\label{incltrends_womass}}
\end{center}
\end{figure*}

\begin{table}
\centering
\caption{Slope values $\gamma_i$ of trends in mass-to-light ratio with inclination: linear fits are obtained for a sample with $\log a/b <0.6$ and a sample without limits. The IR is only available for the higher-mass sample.}
\label{incl_slopes} 
\begin{tabular}{l|cc|cc|}
\hline \noalign{\smallskip}  
			& \multicolumn{4}{|c|}{higher-mass sample}  \\
			& \multicolumn{2}{|c|}{$\log a/b < 0.6$} & \multicolumn{2}{|c|}{all inclinations} \\
passband		& $\gamma_i$ 		& rms 				& $\gamma_i$ 		& rms 	    \\
\noalign{\smallskip} \hline \noalign{\smallskip}
UV 280		& $0.729 \pm 0.229$ & $0.32$ & $0.632 \pm 0.180$ & $0.32$   \\
H$\alpha$		& $0.568 \pm 0.258$ & $0.36$ & $0.503 \pm 0.204$ & $0.36$  \\
IR 24$\mu$	& $0.313 \pm 0.194$ & $0.27$ & $0.119 \pm 0.160$ & $0.28$ \\
\noalign{\smallskip} \hline
			& \multicolumn{4}{|c|}{low-mass sample}  \\
			& \multicolumn{2}{|c|}{$\log a/b < 0.6$} & \multicolumn{2}{|c|}{all inclinations} \\
passband		& $\gamma_i$ 		& rms 				& $\gamma_i$ 		& rms 	    \\
\noalign{\smallskip} \hline \noalign{\smallskip}
UV 280		& $0.101 \pm 0.171$ & $0.32$ & $0.070 \pm 0.115$ & $0.31$ \\
H$\alpha$		& $0.131 \pm 0.188$ & $0.35$ & $0.036 \pm 0.126$ & $0.34$ \\
\noalign{\smallskip} \hline
\end{tabular}
\end{table}



\begin{figure*}
\begin{center}
\includegraphics[angle=270,width=0.95\textwidth,clip=true]{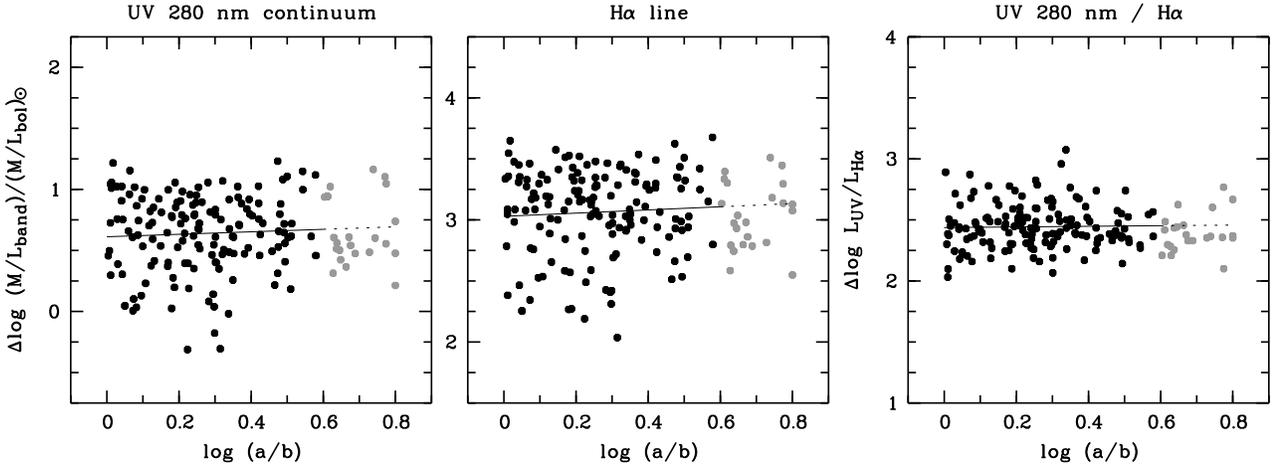} 
\caption{Trends of mass-to-light ratio with galaxy inclination at low mass of $\log M_*/M_\odot =[8.75,9.75]$ after removing a trend with mass, in passbands UV 280~nm continuum (left) and H$\alpha$-emission line (centre), followed by UV-to-H$\alpha$ ratio (right). No significant trends are found. Dark symbols and fits are inclination-restricted; light symbols are highly inclined galaxies, which are not used in the fit but match its extrapolation.
\label{incltrends_lowmass}}
\end{center}
\end{figure*}

\begin{figure*}
\begin{center}
\includegraphics[angle=270,width=0.95\textwidth,clip=true]{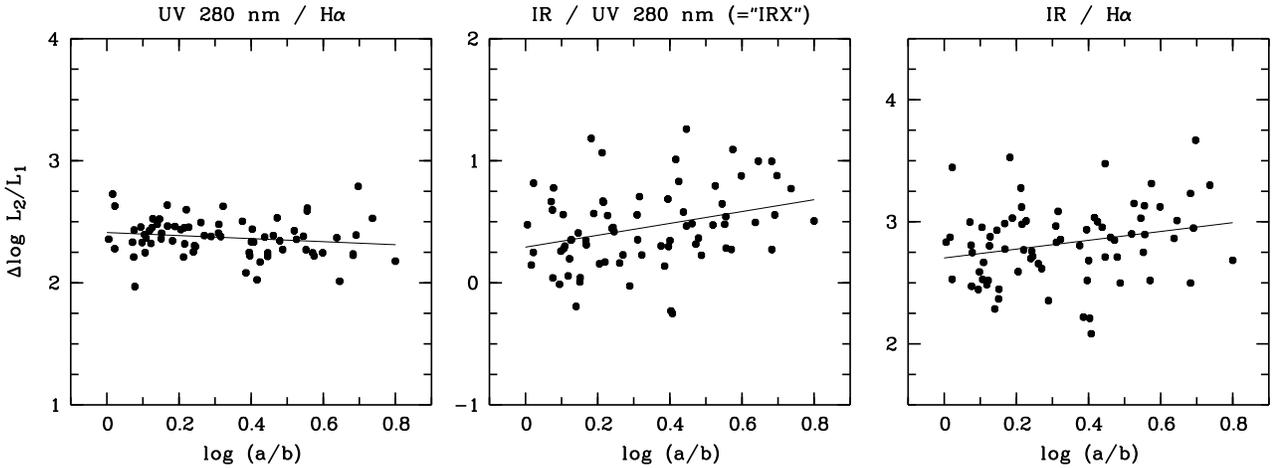} 
\caption{Trends of band-to-band ratio with galaxy inclination: ratio of UV continuum to H$\alpha$-line emission (left), IR to UV continuum ({\it a.k.a.} IRX, centre) and of IR to H$\alpha$-line emission (right). Ratios per galaxy reduce scatter from the variety in star-formation rate among galaxies and show more robust trends by revealing extinction levels per galaxy. In particular, the UV continuum and H$\alpha$-line emission are well-correlated and reveal a modest trend. 
\label{inclratios}}
\end{center}
\end{figure*}

\subsection{Inclination dependence in mass-to-light ratios}

Next we determine trends of mass-to-light ratio with inclination in our three passbands, after removing the mass dependence by subtracting $\beta_i\times (\log M_*/M_\odot -10)$. We obtain linear regression fits and list the slopes in Tab.~\ref{incl_slopes}. 

In the higher-mass sample we find a strong trend in UV with $\gamma_{\rm UV} \approx 0.73$ and a significant trend in H$\alpha$ with $\gamma_{\rm H\alpha} \approx 0.57$ (see Fig.~\ref{incltrends_womass}). When we include high-inclination galaxies, the slope declines by about 10 to 15\% due to the underestimated masses of the added galaxies. In the IR continuum band we still find a slope of $\sim 0.3$ at $1.5\sigma$ significance, which disappears  when including high-inclination galaxies.

At first sight, the absence of a significant trend in the IR band is reassuring, since we claimed this to be radiating optically thin and isotropically, such that any inclination trend should be reflecting biases in determining the stellar masses as opposed to intrinsic galaxy properties. We already know that the masses of high-inclination galaxies are underestimated, suggesting that better mass estimates might reveal a positive trend in the $M_*/L_{\rm IR}$ ratio at $>2\sigma$ significance. This is a hint that either the 24$\mu$-emission underlying our $L_{\rm IR}$ measurement is not isotropic but slightly optically thick and reduced, when a galaxy is viewed edge-on, or that stellar masses are biased such that they get overestimated with inclination at $\log a/b < 0.6$, which corresponds to an axis ratio of $4:1$ or a viewing angle of $\sim 75\degr$ for a thin disk.

In the lower-mass sample we find best-fitting slopes of $\sim 0.1$ with uncertainties larger than that, in both UV and H$\alpha$ (see Fig.~\ref{incltrends_lowmass}). The formal slopes have signs in the expected direction with extinction slightly increasing with inclination, but the effects are weak and statistically insignificant. The low-mass sample ranges from approximately the mass of the Small Magellanic Cloud to that of the Large Magellanic Cloud, two satellites to the Milky Way; galaxies of that mass have irregular shapes and unreliable inclination values, but low-mass galaxies appearing with large axis ratios will still be mostly highly inclined and not intrinsically elongated, and so we conclude that there is no evidence for significant dust column densities in galaxies of $\log M_*/M_\odot =[8.75,9.75]$ that would affect their luminosity or measured star-formation rate as a function of inclination. 

\begin{table}
\centering
\caption{Slope values $\gamma_{i/j}$ of inclination trends in band-to-band light ratios: linear fits are obtained for a sample without inclination limits. The IR is only available for the higher-mass sample. }
\label{incl_ratios} 
\begin{tabular}{l|cc|cc|}
\hline \noalign{\smallskip}  
			& \multicolumn{2}{|c|}{higher-mass sample} & \multicolumn{2}{|c|}{low-mass sample}  \\
passband		& $\gamma_{i/j}$ 		& rms 			& $\gamma_{i/j}$ 		& rms 	    \\
\noalign{\smallskip} \hline \noalign{\smallskip}
IR/UV		& $+0.485 \pm 0.176$ & $0.31$ &  \\
IR/H$\alpha$	& $+0.361 \pm 0.174$ & $0.31$ &  \\
UV/H$\alpha$	& $-0.124 \pm 0.087$ & $0.15$ & $-0.037 \pm 0.066$ & $0.18$ \\
\noalign{\smallskip} \hline
\end{tabular}
\end{table}

\subsection{Extinction changes from face-on to edge-on orientation}

We could now directly estimate how much dust extinction changes from face-on to edge-on geometry in the UV- and H$\alpha$-passbands, but we still intend to refine the fits and remove a potential bias in the $M_*/L_V$ estimates. The trend with inclination in $M_*/L_{\rm IR}$ is less than $2\sigma$-significant, but it is still justified to eliminate this slope, and indeed the scatter introduced by the second nuisance parameter, the star-formation rate, by considering per-galaxy light-to-light ratios from a pair of passbands. Assuming that, at a given stellar mass, the IR-luminosity is a reliable measure of SFR, then the ratio of IR-to-UV or IR-to-H$\alpha$ luminosity reflects the UV- and H$\alpha$-specific extinction levels per galaxy. These ratios may have different normalisations at different stellar mass of a galaxy, but such trends will only increase scatter when plotted against inclination and not bias any trends. Here, we use the whole range of inclinations as the underestimation of stellar masses at high inclinations has no influence.

The results are shown in Fig~\ref{inclratios} and listed in Tab.~\ref{incl_ratios}. The slopes of the light-to-light ratios in a pair are consistent with the differences of the slopes in the two $M_*/L_{\rm band}$ ratios involved, as is of course expected. So, there is a noticeable trend with inclination in the IR-to-H$\alpha$ ratio and a clear trend in the IR-to-UV ratio, a variant of which is also known as IRX or infrared excess \citep{Meurer99,Grasha13}. The uncertainties and scatter are smaller than those of the $M_*/L$ trends themselves. In particular, we see from the UV-to-H$\alpha$ ratio that these two bandpasses are strongly correlated on a per-galaxy basis, constraining the trend to $\gamma_{\rm UV/H\alpha} \approx -0.124$ with a small error of $\sim 0.08$~dex. As expected, the lower-mass sample shows no noticeable trend in the UV-to-H$\alpha$ ratio, with a formal slope estimate of $\gamma_{\rm UV/H\alpha , lowM} = -0.037 \pm 0.066$ (see also right panel in Fig.~\ref{incltrends_lowmass}).

If the low-significance $M_*/L_{\rm IR}$  trend with inclination in the IR is introduced by an inclination bias affecting the mass-to-light ratio from optical broad-band photometry, then we can now determine how much the dust column density increases in the UV and H$\alpha$ passbands as an average disk galaxy turns from face-on to edge-on orientation. We take the inclination range from face-on to edge-on to be given by $\Delta \log a/b = 0.8$ and use the slopes $\gamma_{\rm IR/UV}$ and $\gamma_{\rm IR/H\alpha}$ to determine the extra factor of UV- and H$\alpha$-light that is extinguished by the edge-on line-of-sight. In units of magnitudes, the additional extinction is then
\begin{align}
   \Delta_{0.8} A_{\rm band}  & = 2.5\times 0.8\times \gamma_{\rm IR/band}  \\
   \Delta_{0.8} A_{\rm UV ~ ~ } 	& = 0.97 \pm 0.35 \\
   \Delta_{0.8} A_{\rm H\alpha ~ ~ } & = 0.72 \pm 0.35  ~.
\end{align}  
  
Assuming a \citet{F99} extinction law with $R_V=3.1$, we find $R_{\rm UV}=6.02$ and $R_{\rm H\alpha}=2.36$ and derive the additional reddening due to inclination to be $\Delta_{0.8} E(B-V)_{\rm UVstars} = 0.16 \pm 0.06$ for the UV-emitting stars and $\Delta_{0.8} E(B-V)_{\rm H\alpha gas} = 0.3 \pm 0.15$ for the H$\alpha$-emitting gas regions. This means that we find the diffuse dust column shielding H$\alpha$-emitting gas to be twice as large as that shielding the UV-emitting stellar population. While the large errors on the IR-ratio trends suggest a low significance, the better determined trend of the UV-to-H$\alpha$ ratio reduces the error on the $\Delta_{0.8} E(B-V)$ difference between the two populations to 0.07 and makes it a $2\sigma$-difference. 

These numbers do not represent the full extinction acting on UV-emitting stars and H$\alpha$-emitting gas, because they do not include the face-on part of extinction. It is assumed that especially the H$\alpha$-emitting gas regions are embedded in clumpy dense local dust clouds \citep{Kennicutt83,Tuffs04}, which could well be locally isotropic on average. We assume that the inclination dependence stems from a diffuse dust component in the galaxy disk and that only this component leads to deeper optical depth from longer lines-of-sight when a galaxy is inclined.

\section{Discussion}\label{discussion}

\subsection{Stellar-mass estimates}

A study of inclination effects based on mass-to-light ratios depends crucially on their freedom from inclination biases. The literature contains conflicting statements on the reliability of mass-to-light ratios from \citet{BdJ01}: \citet{Maller09} consider their masses to be the least biased among a choice of three alternatives, and describe them as essentially unbiased even at the highest inclinations; in contrast, \citet{Driver07} consider them to be unbiased at $i<60\degr$, but find them to be biased at higher inclinations. Our results appear in between, as we find these masses acceptable up to $\log a/b=0.6$, which is $i>75\degr$, or indeed $i\approx 80\degr$ for an intrinsic thickness of $q=0.18$. At higher inclinations, all the way to the edge-on case at $i=90\degr$, we find an underestimation of the mass by potentially up to a factor of $\sim 2$. 

This underestimation can be explained by attenuation levels that increase faster than the reddening at the highest inclinations, such that the declining apparent brightness of the galaxy is not sufficiently compensated by an increased mass-to-light ratio. {\refbf \citet{Xilouris99} noted that dust disks have larger scale lengths than stellar disks in galaxies and may extend beyond the edges of the visible stellar disks; if that were the case, such dust would not contribute while inclination changes from face-on to $i\ga80\degr$, but its effect, although presumably weak, could kick in edge-on, when it is moved into the lines-of-sight to the thick disk of a galaxy.}


\subsection{Attenuation of stellar continuum and line-emitting gas}

Our measurement for the attenuation of the UV stellar continuum at 280~nm from face-on to edge-on can be best compared with the results in the SDSS $u$-band by \citet{M10} and in the GALEX FUV-band by \citet{Leslie18}: \citet{M10} found an extinction contribution going from face-on to edge-on of $\Delta A_u=0.7$ in the SDSS $u$-band. Our value of $\Delta A_{280}=0.97$ translates into $\Delta A_u = 0.68$, if we assume a Milky Way extinction law with $R_V=3.1$. We note that \citet{M10} find an attenuation curve, going from $\Delta A_u=0.7$ to $\Delta A_i=0.4$, that is greyer than a Milky Way extinction law, but still consistent with $\Delta A_{280}\approx 1$ within the errors. The results are equally compatible with the slope measurements for the SDSS $u$-band by \citet{Devour16}. \citet{Leslie18} find $\gamma_{\rm FUV}=0.79\pm 0.09$ with a different parametrisation of inclination, suggesting $\Delta A_{\rm FUV}=1.97\pm 0.2$, where we would expect $\Delta A_{\rm FUV}=1.29\pm 0.45$ if we assume a \citet{F99} attenuation curve with $R_{\rm FUV}=8.02$.

In contrast, our results are {\refbf incompatible with the trends suggested by \citet{Maller09}. On this matter, \citet{Devour16} present an excellent discussion suggesting that all} our studies are consistent in terms of data, but lead to inconsistencies due to the analysis, and we have nothing to add to their assessment.

Our value for the attenuation from face-on to edge-on in H$\alpha$ is $\Delta A_{{\rm H\alpha}}=0.72$. This means that the 280~nm-emitting stars are attenuated by less diffuse dust than the Balmer line-emitting gas, most likely with a factor of $\Delta E(B-V)_{\rm stars} \approx 0.52 \times \Delta E(B-V)_{\rm H\alpha}$, but the deviation from equality is only significant at the 2$\sigma$ level. We note, that for the overall extinction ratio \citet{Calzetti00} report $E(B-V)_{\rm stars} = 0.41 \times E(B-V)_{\rm H\alpha}$ in star-bursting galaxies, but this combines the diffuse and the clumpy dust and refers to the population of stars more generally. 

In a simplistic two-component model with a thin disk of diffuse dust and isotropic clumps of thick dust around star-forming regions, we would naively expect no difference in the $\Delta E(B-V)$ of stars and gas caused by inclination, although the face-on $E(B-V)$ values for the two would be very different. This result is indeed found by \citet{Yip10}, whose Fig.~16 shows no inclination trend in the ratio of the equivalent width between H$\alpha$ and H$\beta$ within the errors; this is consistent with the stellar continuum changing with inclination by the same attenuation curve as the Balmer emission lines as this will keep line equivalent widths unchanged\footnote{Note, that the discussion in \citet{Yip10} is misled by a confusion of line fluxes and equivalent widths in their equation 10 for the optical depth; consequently their Fig. 18 is mislabelled on the y-axis as $\tau$ is not really an optical depth, and the diagonal line for the model of dust optical depth being identical for stars and gas should be a horizontal line for their chosen y-axis instead, which would also match their data beautifully. As a consequence, the authors fail to acknowledge that they actually measure a consistent trend in the attenuation of stars and gas with inclination.}.

Still, a marginally higher extinction for the gas could be plausible in a mixed gas-dust-star geometry \citep[and would be consistent with][]{Yip10}, where more stars than gas are seen with low or no attenuation; this would be the case, if either gas is in a thinner disk than the stars and some stars are above the dusty volume and thus less attenuated. An alternative explanation stems from the nature of our cluster sample, which includes some galaxies with more centrally concentrated star formation than is typical in field samples. The loss of star formation in the outer parts of the disks might have gone hand in hand with a loss of gas and dust, such that the outer stellar disks in our galaxies are actually seen with lower-than-typical attenuation that is mixed into our integral values representing galaxies as a whole.

Inclination trends in the attenuation of the line-emitting gas have also been measured by \citet{Battisti17}, who find a mild inclination trend in Balmer decrement, but they consider mostly low-mass galaxies ($\log M_*/M_\odot<10$), and study only the central regions of galaxies through fibre spectra. Interpreting and comparing their results is challenged by the fact that the dust line-of sight in fibre spectra increases with inclination not only by $1/(\cos i)$, but by more as the fibre footprint on the galaxy becomes elliptical and finally cylindrical within the galaxy plane once an edge-on view is reached: thus face-on only the dust within the fibre aperture contributes, while towards edge-on a whole cylinder across the full diameter of the galaxy moves into the fibre aperture.

\subsection{Trends in far-infrared emission}

In this paper we use the far-infrared luminosity as a reference measurement that we expect to be minimally affected by inclination. We find indeed no significant trend in the $M_*/L_{\rm IR}$-ratio with inclination. The same is found by \citet{Leslie18}, who studied WISE W4 data for low-redshift field galaxies in the SDSS.

Due to the restricted dynamic range in mass we find only a weak trend of the $M_*/L_{\rm IR}$-ratio with mass, which is furthermore dominated by a few massive galaxies that have high ratios above trend in all passbands we consider. These are likely to have their masses of $\log M_*/M_\odot\approx 11$ estimated correctly, while having unusually low luminosities in the star-formation indicators because they {\refbf have moved off the star-forming main sequence already}. 

\citet{M10} and \citet{Devour16} have used a much larger parent sample, which allowed a careful analysis of the highest-mass end of the disk galaxy distribution at $\log M_*/M_\odot\approx 11$. They found that the declining star formation is accompanied also by a declining dust density, although this is inferred from the visual extinction and not the infrared emision. The reason is most likely that dust gets produced in the context of star formation, but it also gets destroyed rapidly afterwards and disappears with time if not replenished by sufficient star formation. However, discussions of the life of dust are beyond the scope of this paper.

According to \citet{Calzetti07} and \citet{Calzetti13} a very good star-formation indicator is the sum of star formation seen in H$\alpha$ and 24$\mu$, as it captures both unobscured star formation and dust-reprocessed light from obscured star formation. In the context of our inclination study, we note that such an indicator still shows an inclination trend, as the isotropically emitting 24$\mu$-band is combined with the inclination-dependent H$\alpha$. While the combined indicator is intended to be a best-case estimate in an ensemble average, it will of course still overestimate individual star-formation rates in face-on galaxies and underestimate them at high inclination. 

\subsection{Trends of dust extinction with mass}

{\refbf \citet{Brinch04} and \citet{GarnBest10} measured the mass-dependency of H$\alpha$ extinction for SDSS galaxies. Unfortunately, we are not able to measure trends with mass independently, e.g. using our $24\mu$ data, because the luminosity at $24\mu$ depends not only on star-formation rate but also on mass itself, as evidenced by non-linear calibrations for star-formation rate as a function of 24$\mu$ luminosity; this non-linearity has been studied in great detail and probably results from a mass dependence of metallicity and dust-star geometry \citep[e.g.][]{Calzetti07,Davies16,Brown17}. If we use a mean calibration from \citet{Brown17} to convert 24$\mu$-luminosity into predicted dust-corrected H$\alpha$-luminosity, and derive extinction by comparison with the observed H$\alpha$-luminosity, then we find results compatible with those of \citet{Brinch04} and marginally lower than \citet{GarnBest10}, but the uncertainties are too large to be conclusive.} 



\subsection{Impact of the galaxy cluster on the results}

A valid question is, whether the results we find for our sample of cluster galaxies resembles properties of field galaxies or not. \citet{Tully98} have used cluster samples before, but their disk galaxies from the relatively low-density Ursa Major and Pisces Clusters are expected to be similar to field spirals. The properties of Abell 901/2, our cluster, are more similar to those of the Virgo Cluster than Ursa Major and Pisces; indeed, \citet{Boesch13} found that in A901/2 many of the red spirals in particular have more centrally concentrated and more asymmetric emission-line disks than field spirals. Differences in the radial distribution of young stars, compared to field spirals, propagate into differences in the integrated effect of dust and inclination on star-formation indicators. Our sample contains roughly half-and-half galaxies that are inside and outside of the projected virial radius of the four sub-clusters in the Abell 901/2 system \citep[see e.g.][]{Heymans08,OMEGA3}, but projection effects alone will wash out clear signatures. 

Qualitatively, we expect that ram-pressure stripping or thermal evaporation could have removed some gas and dust from the outer disks of galaxies, especially those that are now seen as less star-forming as a consequence. This effect has been shown to be present in A901/2 by \citet{Boesch13}, and the affected galaxies would now show less average extinction on their stellar population. How the persisting, more centrally located star-forming regions are extinguished in comparison to normal galaxies is not obvious. Central attenuation in normal galaxies is known to be higher than attenuation in outer disks, but ram pressure could have changed the dust-star geometry and models are not refined enough to reliably investigate populations of transitioning galaxies that are not in a known equilibrium configuration.

\section{Summary}

We have studied the effect of inclination on the apparent brightness of normally star-forming galaxies in three spectral passbands commonly used as star-formation indicators. As diagnostics we used mass-to-light ratios in the passbands: UV continuum at 280~nm, the H$\alpha$ emission line, and the far-infrared continuum in the {\it Spitzer} 24$\mu$-band. Our dataset is modest in size, with a few hundred galaxies, but {\refbf novel in its quality: compared to the SDSS, the imaging data are extremely deep in surface-brightness sensitivity, and inclination is measured from the outer disk contours in deep, high-resolution images observed with the {\it Hubble Space Telescope}.} Crucially, the H$\alpha$ emission line is measured via tunable-filter imaging and encompasses galaxies in their entirety, as opposed to fibre spectra that obtain only central measurements at low redshift. The sample should have no selection biases, except that edge-on galaxies with low star formation are under-represented because of the requirements for H$\alpha$-detection. 

Our sample is drawn from a {\refbf volume centred on a few galaxy clusters, and hence the properties of some} galaxies might be affected by the cluster evolution. Differences between these galaxies and field samples are difficult to quantify, but the most likely difference is that our sample includes galaxies with reduced star formation in their outer disks. 

We split our sample into a low-mass bin at $\log M_*/M_\odot = [8.75,9.75]$, which is only complete in UV and H$\alpha$ detections, and a high-mass bin at $\log M_*/M_\odot > 9.75$ that is complete in all three bands for $L_{\rm IR}/L_\odot \ga 9.4$. We find no indication of significant biases in the stellar masses at inclinations up to $i\la 80\degr$, {\refbf except for edge-on galaxies in the higher-mass bin of $\log M_*/M_\odot > 9.75$, which appear underestimated; hence, we ignore this inclination range} where mass enters the analysis. 

In the higher-mass mass sample we find clear inclination trends of the mass-to-light ratio in the UV continuum from young stars and the H$\alpha$ emission line from ionised gas regions, suggesting the presence of diffuse dust in the galaxy disks, which leads to increasing optical depths with increasing inclination. We estimate a dimming of the average massive disk galaxy from face-on to edge-on by $\sim 1$~mag in the 280~nm passband and $\sim 0.7$~mag in H$\alpha$. This implies that star-formation rates measured in UV 280 and H$\alpha$ would be underestimated in nearly edge-on galaxies by a factor of 2.5 and 2, respectively. We find no trend in the IR continuum, suggesting that dust emission is largely optically thin and isotropic in our sample. 

We find a marginal possibility that stars emitting UV continuum at 280~nm may be seen through lower optical depth in diffuse dust than H$\alpha$-emitting gas, but the significance is only $2\sigma$. This finding is irrespective of differences in mean or face-on extinction of the two populations. While it is well-known that H$\alpha$-emitting gas is typically more obscured than stars, our paper concerns only dust effects that manifest themselves with changing inclination, and hence describe the diffuse dust in a two-component model that also includes high-opacity dust clumps around star-forming regions.

The low-mass sample shows no measurable dependence in the brightness of a galaxy on inclination, which is consistent with the view that normally star-forming galaxies with $<5$~billion solar masses have typically low column densities in diffuse dust, such that they are mostly optically thin. They may still have clumpy dust at high column density associated with their star-forming regions, but such dust is expected to have a more locally isotropic distribution, which does not lead to inclination-dependent effects and thus cannot be measured with the approach in this paper.

In a forthcoming paper we will study trends in the total extinction of Balmer lines in the Abell 901/2 sample, using data from the PAnoramic Paschen-Alpha Survey (PAPAS, Wolf et al., in preparation). That work will use the ratio of the H$\alpha$ and the Paschen-$\alpha$ line, and thus measure total extinction and its dependence on mass, inclination and star-formation rate.

\section*{Acknowledgements}
CW acknowledges support from the ARC Laureate Fellowship FL0992131. AAS acknowledges the support from the Australian Research Council Centre of Excellence for All-sky Astrophysics (CAASTRO), through project number CE110001020, for a visit during which part of this paper was written. KH acknowledges the Royal Astronomical Society for financial support through an undergraduate research bursary.
Based on observations made with the Gran Telescopio Canarias, installed in the Observatorio del Roque de los Muchachos of the Instituto de Astrof\'isica de Canarias, in the island of La Palma. The GTC reference for this programme is GTC2002-12ESO. Access to GTC was obtained through ESO Large Programme 188.A-2002. B.R.P acknowledges financial support from the Spanish Ministry of Economy and Competitiveness through grant ESP2015-68964.

\end{document}